\documentclass[prd,aps,twocolumn,amsfonts,showpacs,superscriptaddress,preprintnumbers,nofootinbib]{revtex4-2}

\usepackage{graphicx}
\usepackage{dcolumn}
\usepackage{graphicx}
\usepackage{xcolor}
\usepackage{rotating}
\usepackage{bm}
\usepackage[utf8]{inputenc}
\usepackage{physics}
\usepackage{hyperref}
\usepackage{colortbl}
\definecolor{ddred}{RGB}{198,31,31}
\definecolor{ddgreen}{RGB}{0,160,77}
\hypersetup{
    colorlinks=true,
    linkcolor=ddred,     
    urlcolor=ddred,
    citecolor=ddgreen}

\def\CH{\mathcal{H}}

\usepackage{siunitx}
\makeatletter
\skip\footins=12pt plus 4pt minus 4pt   
\def\footnoterule{%
  \dimen@\skip\footins \divide\dimen@\f@ur   
  \kern-\dimen@\hrule width.5in\kern\dimen@
}
\makeatother

\begin{document}

\title{The Spectral Anatomy of Complex-Time Singularities}

\author{Samuel Valach}
\affiliation{%
 Faculty of Mathematics and Physics, University of Ljubljana, SI-1000, Ljubljana, Slovenia}%

\author{Mile Vrbica}
\affiliation{
 Institute for Theoretical Physics, University of Amsterdam, 1090 GL Amsterdam, The Netherlands
}%
\affiliation{
 Dutch Institute for Emergent Phenomena, 1090 GL Amsterdam, The Netherlands
}%
\affiliation{
 Institute for Advanced Study, University of Amsterdam, Oude Turfmarkt 147, 1012 GC Amsterdam,
The Netherlands
}%

\begin{abstract}
The properties of the bulk spacetime in a holographic theory may be probed by the analytic structure of the boundary correlator $G(t,x)$. In this letter, we analyse the singularities of $G(t,x)$ by deriving them from Fourier space. We show that the quasinormal modes are characterised by two distinct momentum scaling regimes. The first regime is defined by a finite $k/\omega^\gamma$ for some characteristic exponent $\gamma$ which is extracted from the singularity, and probes $G(t,0)$, reproducing the bouncing geodesic analysis. The second regime is defined by a finite $k/\omega$, and probes the complexified bulk away from the singularity, which allows us to study the singularities of $G(t,x\neq 0)$. We extract coefficients and exponents of the singularities for general spacetime dimensions, and allow for more exotic interiors. The heavy reliance of our computation on quasinormal modes opens the doors to a similar analysis being done on spacetimes that are accessible only numerically.
\end{abstract}

\maketitle


\section{Introduction}

\emph{Note: During the final stages of this work, a preprint \cite{Jia:2026qvp} appeared with substantial overlap with our results. The present work was developed independently.} 
\vspace{0.05cm}

In the large-$N$ limit of holographic field theories, the boundary two-point functions $G(t,x)$ encode information about the bulk geometry \cite{Maldacena:1997re, Gubser:1998bc, Witten:1998qj}. Recently, this has attracted a lot of attention in the context of the imprint of the black hole singularity on the boundary. The relevant computation involves a geodesic computation of the analytically extended bulk spacetime, known in this context as the `bouncing geodesic' \cite{Fidkowski:2003nf,Festuccia:2005pi}, which leads to a branch point of $G(t,0)$ in complex time \cite{Parisini:2023nbd,Ceplak:2024bja,Grozdanov:2026cut} -- the bouncing singularity. These studies often require a study of complexified geodesics and OPE \cite{Parisini:2023nbd,Ceplak:2024bja,Buric:2025fye,Valach:2025saf,Barrat:2025twb,Ceplak:2025dds,Jia:2026pmv,Araya:2026shz,Buric:2026qsp,Buric:2026cmn,Arnaudo:2026axe,Barrat:2026jfg,Ceplak:2026ard,Batra:2026gbq,Komargodski:2026oob,Buric:2026sit,Batra:2026otu,Dodelson:2026hjm,Oh:2026dky,Oh:2026qcm,Chakravarty:2026xvr}, and become especially difficult when $x\neq 0$. Another perspective is offered by the Fourier space perspective, by studying $G(\omega,k)$. There, the correlator is determined by the quasinormal modes (QNMs) \cite{Dodelson:2023vrw}, which allows for derivation of the singularities $G(t,k)$ via large-$\omega$ trans-asymptotics \cite{Afkhami-Jeddi:2025wra,Dodelson:2025jff,Jia:2025jbi,AliAhmad:2026wem,Giombi:2026kdz,Grozdanov:2026cut,Jia:2026ryl,Arnaudo:2026tcy,Grozdanov:2026ktq,Grozdanov:2026lnc,Hartnoll:2026vhu,Coviello:2026hdc,Saha:2026ows,Buric:2026sit,Shi:2026cfq,Chakravarty:2026fdm,Chen:2026efa,Arnaudo:2026gtd,Liu:2026luf}. However, fixed-$x$ and fixed-$k$ correlators have different analytic structure in complex time, obscuring the precise connection between the QNM and the geodesic approaches \cite{Grozdanov:2026cut}.
\begin{figure}[ht]
\centering
\includegraphics[width=0.45\textwidth]{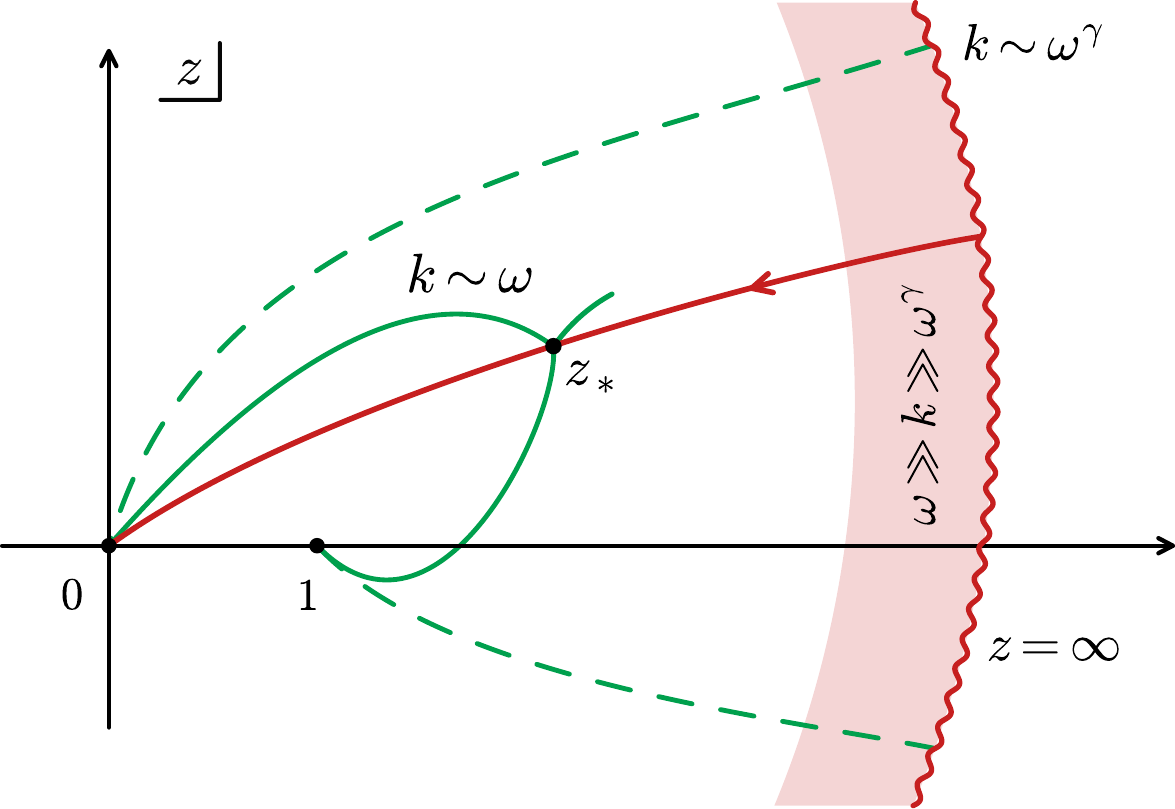}
\caption{Schematic representation of the distinct WKB regimes determining physics in the complex-$z$ plane. Here the boundary is located at $z=0$, horizon at $z=1$ and curvature singularity at $z=\infty$. The green lines track the anti-Stokes lines in the outer-$k$ regime $k\sim\omega$ (solid green) and inner-$k$ regime $k\sim\omega^\gamma$ (dashed green). Slightly away from the singularity the red area denotes the intermediate regime $\omega\gg k\gg\omega^\gamma$. The solid red line tracks the movement of the turning point $z_*$ for the WKB in the direction of growing $k$.}
\label{f.regimes}
\end{figure}

In this work, we address this gap by studying the large-$k$ behaviour of QNMs and performing the subsequent Fourier transform. We are specifically treating a $d$-dimensional boundary ($d\geq 3$) dual to the planar anti-de Sitter-Schwarzschild black hole (AdS-SC), and a probe scalar with scaling dimension $\Delta=d/2+\nu$. Many of the results of this paper are, however, applicable to more general geometries, such as black holes with a Kasner interior \cite{Kasner:1921zz,Belinsky:1970ew,Hartnoll:2020rwq,Frenkel:2020ysx,Hartnoll:2026vhu}. We perform a WKB analysis in large $\omega$ to derive the asymptotic QNMs at various momenta $k$. We derive several distinct momentum regimes, the classification of which is heavily dependent on the near-singularity behaviour of the bulk equations of motion. More specifically, the singularity singles out an exponent $\gamma$ with $0<\gamma<1/2$, and the classification of large-$n$ QNMs is the following:\footnote{Throughout the paper, any scale comparison of $k$ with $\omega$ is to be understood as taking the absolute value of $\omega$, and any scale comparison of $k$ with an integer $n$ is to be understood by endowing $n$ with the scale of temperature.}

\begin{itemize}
    \item The \emph{small}-$k$ \emph{regime} $k \ll n^\gamma $ is the one that has been studied extensively in WKB studies of QNMs \cite{Natario:2004jd,Cardoso:2004up,Motl:2003cd}, with the first correction given in \cite{Hartnoll:2026vhu}:
    \begin{equation}\label{s.regime_SMALL}
        \omega_n \sim \frac{\pi}{\mathcal{Z}}\qty[n+\frac{\nu-1}{2} +i\frac{\log 2}{2\pi}+\alpha_0\frac{k^2}{n^{2\gamma}}].
    \end{equation}
    \item The \emph{inner}-$k$ \emph{regime} $k \sim n^\gamma$ is characterised by a non-trivial function $F$ that must be computed from near-singularity physics
    \begin{equation}\label{s.regime_INNER}
         \omega_n\sim \frac{\pi}{\mathcal{Z}}\qty[n+\frac{\nu-1}{2}+F\qty(\frac{k}{n^\gamma})].
    \end{equation}
    \item The \emph{intermediate-$k$ regime} $n\gg k \gg n^\gamma$ corresponds in the bulk to the region which is close to the singularity in terms of the AdS radius, but far from singularity in terms of the wavelength of the scalar solution
    \begin{equation}\label{s.regime_INTERMEDIATE}
        \omega_n \sim \frac{\pi}{\mathcal{Z}} n \qty[1+\alpha_1 \qty(\frac{k}{n})^{2-\frac{2}{d}}].
    \end{equation}
    \item The \emph{outer-$k$ regime} $k\sim n$ probes the complexified bulk away from the singularity, and is governed by functions $H(k/n)$ and $\bar H(k/n)$ that can be expressed in closed form. Explicitly,
    \begin{equation}\label{s.regime_OUTER}
    \omega_n\sim n \, H(k/n)+\frac{\nu-1}{2}\bar H(k/n).
    \end{equation}
    \item The \emph{near-null regime} $k\gg n$ probes the near-boundary region and was previously derived in \cite{Festuccia:2008zx,Fuini:2016qsc}
    \begin{equation}\label{s.regime_NULL}
        \omega_n \sim k+ \alpha_2\, n \qty(\frac{k}{n})^{-\frac{d-2}{d+2}}.
    \end{equation}
\end{itemize}
The definition and value of the optical distance $\mathcal{Z}$ and explicit expressions for $\alpha_0$, $\alpha_1$, and $\alpha_2$ are provided below. The small-$k$ and inner regimes probe the singularity (see dashed green line on Fig.~\ref{f.regimes}), the outer regime probes the far-from-singularity bulk (see green line on Fig.~\ref{f.regimes}), and the near-null regime probes the near boundary region. The intermediate regime is especially interesting, as it probes the region which is close to the singularity in terms of the AdS radius, but far from it in terms of the characteristic wavelength of the solution $\sim 1/\omega$ (see red region in Fig.~\ref{f.regimes}).

To derive $G(t,x)$ from QNMs, one would require the full momentum dependence, not just the relevant scaling regimes. However, we show that the results presented above are sufficient to study the near singularity behaviour of the retarded correlator
\begin{equation}\label{e.sing}
    G(t\rightarrow t_c,x)\sim \frac{F_c}{(t-t_c)^r}, 
\end{equation}
specifically, to compute the position of the singularity $t_c$, coefficient $F_c$, and the power $r$, all of which are a function of $x$. When $x=0$, those can be derived from the inner scaling regime \eqref{s.regime_INNER}, with the exponent $\gamma$ playing the central role. This is coherent with the bouncing geodesic story, where the expression \eqref{e.sing} descends from the geodesic bounce. When $x\neq 0$, the data in Eq.~\eqref{e.sing} can be derived from the outer regime \eqref{s.regime_OUTER}, probing instead the far-from-singularity region of spacetime. The results presented here agree with and extend the results currently present in literature. Importantly, the analysis presented here may be extended to more interesting bulk geometries as well as geometries that are only accessible numerically. This is due to the fact that QNMs are easily computable on a numerical background, in great contrast to complex geodesics.

In Section~\ref{s.WKBsetup} we outline the WKB computation, which we then apply to the inner, outer and intermediate scaling regimes. In Section~\ref{s.FT} we perform the Fourier computation and connection to both -- bouncing and bulk-cone geodesics. In Section~\ref{s.Discusion} we discuss the application of the computation to more general geometries and potential future directions.


\section{WKB computation}\label{s.WKBsetup}

Throughout, we are working with $\omega$ in the fourth complex quadrant ($\Im\omega<0$ and $\Re\omega>0$), a real positive $k$, and a complexified bulk radial coordinate so that $z=0$ is the boundary, $z=1$ the horizon, and $z=\infty$ the singularity (see Fig.~\ref{f.regimes}). That is encoded in the emblackening factor $f(z)=-z^2 g_{tt}=1/z^2 g_{zz}$ ($f(z)=1-z^d$ in the AdS-SC). It is convenient to introduce
\begin{equation}
    \mu\equiv \frac{k}{\omega},
\end{equation}
and consider the physics away from the lightcone $\abs{\mu}<1$.

We are interested in solving the bulk wave equation of a minimally coupled scalar in the regime of large $\omega$. The WKB solution approximates solutions well when the characteristic wavelength of solution is much smaller than the characteristic scale of the coefficients of the equation of motion. This is precisely applicable to large-$\omega$ limit of the bulk wave equation (see e.g. \cite{Natario:2004jd,Cardoso:2004up,Motl:2003cd,Hartnoll:2026vhu,Jia:2025jbi,Jia:2026ryl,Afkhami-Jeddi:2025wra,Stephanov:2011wf} for the technical details). We introduce
\begin{equation}
    p(z)\equiv\frac{\sqrt{1-\mu^2 f(z)}}{f(z)}.
\end{equation}
The solution is valid away from the zeroes and singularities of $p(z)$. We set the ingoing boundary condition at the horizon, meaning that, in the near-horizon matching region
\begin{equation}
    \psi_\text{horizon}(z)\sim e^{i \omega \int^z p(z')\dd z'}.
\end{equation}
The WKB approximation breaks down near the boundary, where an explicit near-boundary solution must be constructed, relating the boundary fall-offs to the WKB solution in the respective matching region. We introduce a function $\Omega(\omega,k)$ that parametrises the specific superposition of the two near-boundary solutions, so that, in the matching region,
\begin{equation}
    \psi_\text{bdry}(z)\sim i e^{2 \pi i \Omega+i \omega \int^z p(z')\dd z'}+ e^{-i \omega \int^z p(z')\dd z'}.
\end{equation}
The boundary and the horizon WKB solution may be matched to one another by matching the coefficients in front of the oscillatory exponentials. The exponentials are oscillatory only when
\begin{equation}
    \Im \Big(\omega \int^z p(z')\dd z'\Big)=0,
\end{equation}
which defines anti-Stokes lines in the complex $z$ plane. For generic values of $\omega$ and $k$, there is always a trivial solution as there is an anti-Stokes line connecting the horizon and boundary directly, giving $\Omega \rightarrow -i \infty$. The non-trivial solutions relevant for our study come when the anti-Stokes lines emanating from the boundary and horizon meet at a turning point or the singularity, yielding a non-trivial $\Omega(\omega,k)$. The boundary retarded correlator may be expressed as
\begin{equation}
    G(\omega,k)\approx \frac{\pi^\frac{d}{2}\Gamma(-\nu)}{4^\nu \Gamma\qty(\frac{d}{2}+\nu)}(\omega^2-k^2)^\nu \frac{\cos \pi \qty(\Omega(\omega,k)-\frac{\nu}{2})}{\cos\pi \qty(\Omega(\omega,k)+\frac{\nu}{2})}, \label{retarded}
\end{equation}
where the normalisation is chosen so that $G(x)=1/x^{2\Delta}$ in the vacuum. The WKB approximation holds well when $\omega\rightarrow \infty$ in the direction of the complex plane where $\Im \Omega \approx 0$ and $\Im \omega<0$. The quasinormal modes are the solutions of
\begin{equation}
    \Omega(\omega,k)+\frac{\nu-1}{2}+n=0.\label{QNM-condition}
\end{equation}

\subsection{Outer regime (\texorpdfstring{$k\sim \omega$}{k~w})}\label{s.largemomentum}

At finite $\mu$, i.e.~when $k \sim \omega$, the non-trivial solutions to the WKB problem are the ones where there is an anti-Stokes line emanating from the boundary, hitting a turning point $z_*$, and connecting to the horizon by an anti-Stokes line that is rotated for $-2\pi/3$ (see solid green curves in Fig.~\ref{f.regimes}). The relevant turning point is the solution to
\begin{equation}
    z_*^d=1-\frac{1}{\mu^2},
\end{equation}
with $\Re z_*>0$ and $\Im z_*>0$ (in case of more such solutions, the relevant one is the one with the smallest imaginary part). The problem is solved by the standard Airy solution around the turning point. The result of the computation is
\begin{equation}
    \Omega=-\Psi(\mu)\,\omega + \mathcal{O}(\omega^{-1}),
\end{equation}
where
\begin{equation}
    \Psi(\mu)\equiv \frac{1}{\pi}\int_0^{z_*} p(z')\dd z',
\end{equation}
the contour of integration being taken as suggested by the red line in Fig.~\ref{f.regimes}. This integral may be expressed in closed form
\begin{equation}
    \Psi(\mu)=\frac{z_*\sqrt{1\!-\!\mu^2}}{d\pi}B\qty(\tfrac{1}{d},\tfrac{3}{2}){}_2F_1\qty(1,\tfrac{1}{d},\tfrac{3}{2}\!+\!\tfrac{1}{d};1\!-\!\tfrac{1}{\mu^2}).
\end{equation}
Using Eq.~\eqref{QNM-condition}, the QNMs may be expressed by introducing $H(x)$ such that
\begin{equation}
    H(x)\Psi\Big(\frac{x}{H(x)}\Big)=1, \quad \bar H(x)=H(x)-x H'(x).
\end{equation}
The limit of $\mu\rightarrow 1$ yields an expansion 
\begin{equation}
    \Psi(\mu\rightarrow1)\sim (1-\mu^2)^{\frac{1}{2}+\frac{1}{d}}e^{i\frac{\pi}{d}}\frac{B\qty(\frac{1}{d},\frac{3}{2})}{d\pi},
\end{equation}
giving (see Eq.~\eqref{s.regime_NULL})
\begin{equation}
    \alpha_2=\frac{1}{2}\qty[e^{-i\frac{\pi}{d}}\frac{d\pi}{B\qty(\frac{1}{d},\frac{3}{2})}]^\frac{2d}{2+d}.
\end{equation}
The limit of $\mu \rightarrow 0$ gives
\begin{equation}
   \Psi(\mu\rightarrow 0)\sim \frac{1}{\pi}\qty[\mathcal{Z}+\mu^{2-\frac{2}{d}}e^{i\frac{\pi}{d}}\frac{B\qty(\frac{1}{2},\frac{1}{d})}{2(1-d)}],
\end{equation}
where we have introduced the \emph{optical distance} (see e.g. \cite{Hartnoll:2026vhu})
\begin{equation}
    \mathcal{Z}\equiv \pi \Psi(0)=\theta\frac{ e^{i\theta}}{\sin \theta}
\end{equation}
where $\theta=\pi/d$. It is related to the inverse temperature $\beta$ through
\begin{equation}
    \Im\mathcal{Z}=\frac{\beta}{4}.
\end{equation}
This limit will be elaborated on in Section~\ref{s.intermediatemomentum}.

\subsection{Inner regime (\texorpdfstring{$k\sim \omega^\gamma$}{k~w\^gamma})}

When $\mu=0$, we get $p(z)\approx 1/f(z)$, and $z_*$ reaches the singularity and the computation does not involve a turning point anymore. There is an anti-Stokes line emerging from the boundary to the singularity with $\arg(z)\sim \theta$ and another from the horizon to the singularity with $\arg z \sim \theta-\pi/(d-1)$, see green dashed lines in Fig.~\ref{f.regimes}. These lines enter the near-singularity regime where WKB breaks down. We must therefore match the near-singularity solution to the two anti-Stokes lines. To be more concrete, we introduce a new variable $y$
\begin{equation}
    y\equiv \omega\qty(\int_0^z\frac{\dd z'}{f(z')}-\mathcal{Z})
\end{equation}
so that, near the singularity $y\sim \omega z^{1-d}/(d-1)$. The near-singularity equation reduces to
\begin{equation}\label{eq-nearsing}
    \Big(\frac{\dd^2}{\dd y^2}+1+\frac{1}{4y^2}+\frac{\lambda^2}{y^{2(1-\gamma)}}\Big)\psi(y)=0,
\end{equation}
where
\begin{equation}
    \lambda \equiv \sqrt{A} \frac{k}{\omega^\gamma}. \label{def:lambda}
\end{equation}
In the AdS-SC case, we have
\begin{equation}
    A=(d-1)^\frac{d}{1-d}, \quad \gamma=\frac{d-2}{2(d-1)}. \label{eq-gamma}
\end{equation}
This calculation, however, is applicable to more exotic singularities, such as Kasner. We therefore leave $A$ and $\gamma$ arbitrary, so long as $0<\gamma<1/2$. In the matching regions, we connect to the ingoing boundary conditions from the horizon
\begin{equation}
    \psi_\text{sing}(y)\propto e^{iy},\\
\end{equation}
and to the boundary via
\begin{equation}
        \psi_\text{sing}(ye^{-i\pi})\propto ie^{2\pi i \Phi(\lambda)-i y}+e^{i y}
\end{equation}
where the $e^{-i\pi}$ continuation corresponds precisely to matching the solution between the horizon and boundary anti-Stokes line. This introduces $\Phi(\lambda)$ as a function encoding the mapping between the matching coefficients. While there is no explicit form for $\Phi(\lambda)$, it is easily computable by numerical integration. By matching the solutions on the anti-Stokes lines, we get
\begin{equation}
    \Omega=-\frac{\mathcal{Z}}{\pi}\omega + \Phi(\lambda). \label{wkb-phi}
\end{equation}
Solving for the quantisation condition \eqref{QNM-condition} yields
\begin{equation}
    F(x)=\Phi\qty(\frac{\mathcal{Z}^\gamma}{\pi^\gamma}\sqrt{A}x).
\end{equation}
The small-$\lambda$ regime of $\Phi(\lambda)$ has been derived in \cite{Hartnoll:2026vhu}
\begin{equation}
        \Phi(\lambda\rightarrow 0) \sim i\frac{\log{2}}{2 \pi}+\lambda^2 \frac{i e^{-i \pi \gamma}B\qty(\frac{1}{2},\frac{1}{2}-\gamma)}{4\Gamma(1-\gamma)^2}+\ldots,
\end{equation}
giving, in Eq.~\eqref{s.regime_SMALL},
\begin{equation}
    \alpha_0=A \qty(\frac{\mathcal{Z}}{\pi})^{2\gamma}\frac{i e^{-i \pi \gamma}B\qty(\frac{1}{2},\frac{1}{2}-\gamma)}{4\Gamma(1-\gamma)^2}.
\end{equation}

\subsection{Intermediate regime (\texorpdfstring{$\omega \gg k \gg \omega^\gamma$}{w<<k<<w\^gamma})} \label{s.intermediatemomentum}

As mentioned before, the small-$k$ regime \eqref{s.regime_SMALL} and the near-null regime \eqref{s.regime_NULL} correspond to the near-singularity and near-boundary WKB matching computations respectively. The intermediate regime is more interesting and subtle, and may be derived in two distinct, but ultimately equivalent, ways. 

The first way is to expand $\Phi(\lambda)$ for large $\lambda$. This involves performing a WKB analysis on Eq.~\eqref{eq-nearsing}. The anti-Stokes lines at $y\rightarrow \infty^\pm$ are matched via a turning point at
\begin{equation}
    y_*=\qty(\lambda e^{-i \frac{\pi}{2}})^\frac{1}{1-\gamma},
\end{equation}
with a standard Airy computation rotating for $-2\pi/3$ from the boundary line to the horizon line. This gives
\begin{equation}
\Phi(\lambda\rightarrow\infty)\sim\lambda^{\frac{1}{1-\gamma}}\frac{e^{i\pi \frac{1-2\gamma}{2(1-\gamma)}} B\qty(\frac{1}{2},\frac{1-2\gamma}{2(1-\gamma)})}{2
        \pi}. \label{largephi}
\end{equation}
On the other hand, we can expand $\Psi(\mu)$ for small $\mu$, giving
\begin{equation}
        \Psi(\mu\rightarrow 0)\sim \frac{1}{\pi}\qty[\mathcal{Z}+\mu^{2-\frac{2}{d}}e^{i\frac{\pi}{d}}\frac{B\qty(\frac{1}{2},\frac{1}{d})}{2(1-d)}].
    \end{equation}
Taking Eq.~\eqref{eq-gamma} into account, these two expressions contain precisely the same data, and an overlap identity holds:
\begin{equation}
    \lambda \gg 1 \gg \mu: \quad  \frac{\mathcal{Z}}{\pi}-\Psi(\mu)= \frac{\Phi(\lambda)}{\omega}.
\end{equation}
This gives, in Eq.~\eqref{s.regime_INTERMEDIATE},
\begin{equation}
    \alpha_1=-e^{i\frac{\pi}{d}}\frac{B\qty(\frac{1}{2},\frac{1}{d})}{2\pi(1-d)}\qty(\frac{\pi}{\mathcal{Z}})^{\frac{2}{d}-1}.
\end{equation}
The fact that the expansions in two distinct regimes have an overlap is expected, as there is no intermediate scale that could interfere with the smooth transition. Nevertheless, the fact that the two computations give the same result serves as an important validity check, and it is pertinent to understand it better. The outer regime holds whenever the turning point $z_*$ lies many WKB wavelengths away from the singularity. On the other hand, the inner regime holds whenever the matching computation is done in the near-singularity region. The overlapping region is
\begin{equation}
    1\ll \abs{z_*}^{d-1}\ll \abs{\omega},
\end{equation}
which singles out a specific region of the complexified bulk which is close to the singularity in terms of AdS radius, but far away from it in terms of the WKB wavelength (see red region in Fig.~\ref{f.regimes})


\section{Fourier integrals and complex time singularities}\label{s.FT}

To study the singularities of $G(t,x)$, it is convenient to expand the cosine of Eq.~\eqref{retarded} into a geometric series
\begin{equation}\label{e.squirrel}
    G(\omega,k)\propto \sum_n G_n(\omega,k)
\end{equation}
with
\begin{equation}
    G_n(\omega,k)\propto (\omega^2-k^2)^\nu e^{-2i\pi n \qty(\Omega(\omega,k)+\frac{\nu-1}{2})}.
\end{equation}
where the proportionality holds up to the $\Omega$-independent $n=0$ term, which will not affect the outcome of our computation.\footnote{Note also that the manipulations of this section are identical for the retarded correlator or the spectral function $\rho\sim \expval{[\mathcal{O}(t,x),\mathcal{O}(0,0)]}$ as the poles on the upper half plane do not affect the computation.} For now, we discard the universal proportionality factors and recover them for the final results. The crucial observation is the fact that a singularity of the type \eqref{e.sing} emerges from integrals of the type
\begin{equation}
    \int^\infty (\omega^{r-1} +\ldots) e^{i\omega(t_c-t)+\ldots}\,\dd\omega\sim \frac{\Gamma(r)}{(it-it_c)^r}, \label{gammaeq}
\end{equation}
where some convergent contour must be chosen. Non-equivalent choices of the contour for the same values of complex $t_c-t$ correspond to picking the sheet when going around the branch point in complex time. Crucially, the singularity's leading coefficient, exponent, and location do not depend on the subleading terms denoted by the ellipsis \cite{Dodelson:2025jff}, nor on the lower bound of integration. We are therefore justified in using the large-$\omega$ WKB analysis of Section~\ref{s.WKBsetup} to obtain singularity results. Secondly, since $\Omega \sim \omega$, every term in the sum of \eqref{e.squirrel} is expected to result in a singularity $t_n$. Importantly, two terms with a different exponent cannot contribute to each other's leading singularity behaviour. Therefore it is justified to treat the terms in the sum independently.

\subsection{\texorpdfstring{$x=0$}{x=0}}
We consider the integral
\begin{equation}
    G(t,0)\propto \int^\infty e^{-i \omega t}  \dd \omega \int_0^\infty k^{d-2} \dd k \, G(\omega,k).
\end{equation}
We may do a substitution $k=k(\lambda)$ (see Eq.~\eqref{def:lambda}). For large $\omega$ we get
\begin{equation}
\begin{split}
     k^{d-2} &\dd k \, G(\omega,k) \sim  \omega^{2\nu+\gamma(d-1)}A^\frac{1-d}{2} \cross \\
     &\hspace{1.5cm}\cross \sum_n  \lambda^{d-2}\, e^{-2\pi i n\qty(\Omega+\frac{\nu-1}{2})}\, \dd \lambda.
\end{split}
\end{equation}
If we pick $\arg{\omega}=-\theta$, we may plug in the WKB result of Eq.~\eqref{wkb-phi}. The leading large $\omega$ behaviour factorises into the $\omega$-dependent and the $\lambda$-dependent factors, meaning we can integrate over them separately.\footnote{This would not have happened for a substitution $k=u \omega^s$ for any $s\neq \gamma$.} The $x=0$ computation therefore maps onto the inner regime, which is exactly what we would expect from the bouncing geodesic picture. We define
\begin{equation}
\label{def:M}
    M(n)\equiv \int_0^{e^{i\gamma\theta}\infty} \lambda^{d-2}\dd \lambda \,e^{-2\pi i n \qty(\Phi(\lambda)+\frac{\nu-1}{2})}.
\end{equation}
Due to \eqref{largephi}, the integral is not convergent for positive $n$, which is a consequence of the geometric series expansion not commuting with the momentum integral. We must therefore evaluate $M(n)$ in some convergent region and then analytically continue.\footnote{Inequivalent analytic continuation may take us on different sheets of complex time.} This may be justified by a careful resummation. Note that the integral \eqref{def:M} is evaluated strictly for the complex argument that emerges in the QNM expression \eqref{s.regime_INNER}, and is therefore directly accessible from quasinormal modes.

Defining $t_n=2n \mathcal{Z}$ and using Eq.~\eqref{gammaeq}, the $\omega$ integral yields
\begin{equation}\label{e.finaler}
    G(t\rightarrow t_n,0)\sim C\frac{A^\frac{1-d}{2} M(n)\Gamma(p) e^{i\pi \frac{1-p}{2}}  }{(t-t_n(0))^p},
\end{equation}
where $p=2\nu+1+(d-1)\gamma$, and
\begin{equation}
    C=\frac{ 2^{2-d-2\nu}\sqrt{\pi}}{\Gamma\qty(\frac{d-1}{2})\Gamma(1+\nu)\Gamma\qty(\frac{d}{2}+\nu)}.
\end{equation}
We emphasise again that the normalisation is chosen so that $G(x)=1/x^{2\Delta}$ in the vacuum. The location of the singularities $t_n$ recovers the standard bouncing geodesic result at $n=1$ \cite{Ceplak:2024bja}, including the correct exponents, and extends the results to other values of $n$. These should be understood as $n$-bounces off the curvature singularity, see e.g.~\cite{Afkhami-Jeddi:2025wra,Dodelson:2025jff,Grozdanov:2026cut}. 

\subsection{\texorpdfstring{$x\neq 0$}{nonzero x}}

We now consider $x\neq 0$. Similarly as for the small momentum, we Fourier transform each individual term $G_n(\omega,k)$ in the sum \eqref{e.squirrel}, which, after substitution $k=\mu\omega$, lead to
\begin{equation}\label{e.stpt1}
    G_n(t,x)\propto x^\frac{3-d}{2} \!\int^\infty\!\!e^{-i \omega t}\dd \omega\, \CH_n(\omega),
\end{equation}
where
\begin{equation}\label{eq:Hn-u}
\CH_n(\omega)=\omega^{\frac{d+1}2+2\nu}\!\!\int_0^\infty \!\!\!\dd\mu\,\mu^{\frac{d-1}2}J_{\frac{d-3}2}(x\omega \mu)\,(1-\mu^2)^\nu e^{2\pi i n\omega\Psi(\mu)}\!.
\end{equation}
By arguments similar to before, the approximation here is justified by the fact that it isolates the leading large-$\omega$ behaviour. Expanding the Bessel function in the large argument, we get
\begin{equation}\label{eq:Hn-split}
\begin{split}
    \CH_n(\omega)\sim\,&\frac{\omega^{\frac d2+2\nu}}{\sqrt{2\pi x}}\sum_{\pm}\!e^{\mp i\frac\pi4(d-2)}\, \times\\
    &\times \int\!\dd\mu\,\mu^{\frac{d-2}2}\!(1-\mu^2)^\nu e^{i\omega(\pm x \mu+2\pi n \Psi(\mu))}.
\end{split}
\end{equation}
This integral can now be treated using the saddle point approximation; the saddle points $\mu_\pm$ are located at
\begin{equation}
    \pm x+2\pi n \Psi'(\mu_\pm)=0,
\end{equation}
giving an explicit expression
\begin{equation}\label{eq:saddle}
    x=\pm\frac{2n}{d}\,e^{i\pi/d}\,B(\tfrac12,\tfrac1d)\qty[\frac{\mu_\pm}{\sqrt{1-\mu_\pm^2}}]^{\frac{d-2}{d}} .
\end{equation}
The complex time singularities then appear at
\begin{equation}\label{eq:tn}
    t_n^+(x)=x\mu_++2\pi n\Psi(\mu_+),\qquad t_n^-(x)=t_n^+(-x),
\end{equation}
where $x$ is implicitly expressed by $\mu_\pm$. Evaluating Eq.~\eqref{eq:Hn-split} on the saddles, we get
\begin{align}
    \CH_n(\omega)&\sim x^{-1/2}\sum_{\pm}F_n^\pm(x)\,\omega^{q-1}e^{i\omega t_n^\pm(x)},\label{eq:Hn-final}\\
    F_n^\pm(x)&=\frac{e^{\mp i\frac\pi4(d-2)}\,\mu_\pm^{\frac{d-2}2}(1-\mu_\pm^2)^\nu}{\sqrt{\frac{2in(d-2)}{d^2}\,e^{i\frac{\pi}{d}}B(\frac12,\frac1d)\mu_\pm^{-\frac2d}(1\!-\!\mu_\pm^2)^{\frac1d-\frac32}}},
\end{align}
where we assumed both saddles are crossed by the deformed contours. Finally, plugging this into \eqref{e.stpt1}, and decomposing we get the final result as
\begin{equation}
\label{eq:mainrez}
    G(t\rightarrow t_n^\pm(x),x)\sim D \frac{x^{1-\frac d2}F_n^\pm(x) e^{i\pi(n(1-\nu)-\frac{q}{2})}\Gamma(q)}{\big(t-t_n^\pm(x)\big)^{q}},
\end{equation}
where $q=2\nu+\frac{d+1}2$ and
\begin{equation}
    D=i\frac{ \sqrt{\pi} 2^{\frac{1-d}{2}-2\nu}}{\,\Gamma(1+\nu)\,\Gamma\big(\frac d2+\nu\big)}.
\end{equation}

\begin{figure}[ht]
\centering
\includegraphics[width=0.36\textwidth]{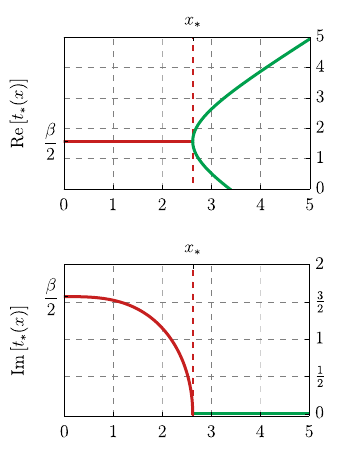}
\caption{Behaviour of the saddle time $t_*$ \eqref{eq:d4real} and \eqref{eq:d4complex} as a function of $x\in\mathbb{R}$ in $d=4$. For $x<x_*$ the time $t_*$ smoothly connects the standard $x=0$ bouncing singularity and the junction point. For $x>x_*$ the time $t_*$ continuously transits to the (complex) bulk-cone singularity of \cite{Araya:2026shz}. For $x\to0$, $\text{Im}(t_*)$ obeys the expected $x^3$ behaviour.}
\label{f.adam}
\end{figure}

First, note that the power $q$ of the singularity differs from the case $x=0$ by $q-p=1/2$. This result precisely matches one of the complex bulk-cone singularities, which are expected to describe the large-$x$ regime, cf.~Eq.(4.27) of \cite{Araya:2026shz}. As a further comparison, taking $\text{Im}(x)=0$, one reconstructs the location of the complex bulk-cone singularity \cite{Araya:2026shz} and continuously connects it to the bouncing singularity, simply using the saddle point result \eqref{eq:tn}. Concretely, taking \eqref{eq:tn}, defining $t_*=t_1^+$ and focusing on $d=4$, one finds these two regimes to be connected by a junction point at $x_*=B(\tfrac12,\tfrac14)/2\approx2.6221$ where the behaviour sharply, yet continuously, changes (see Fig.~\ref{f.adam}) 
\begin{align}
    0\le x\le x_*&:\quad\! t_*(x)=\frac\beta2+i\,\frac\beta2\Big[1-I_{\tfrac{x^4}{x_*^4}}\big(\tfrac34,\tfrac12\big)\Big],\label{eq:d4complex}\\
    x\ge x_*&:\quad\! t_*(x)=\frac\beta2\pm\frac{B(\tfrac12,\tfrac14)}{4}\!\int_1^{\frac{x^2}{x_*^2}}\!\!\frac{\sqrt{y}\,\dd y}{\sqrt{y^2-1}},\label{eq:d4real}
\end{align}
where $I_{z}(a,b)$ is the regularised incomplete Beta function. Interestingly, the existence of such junction point is special to $d=4$.

Finally, let us examine the small-$x$ limit of these results. Expanding $t^+_n$ for $x\to0$, we find
\begin{equation}\label{eq:smallx}
    t_n^+(x)-2n\mathcal{Z}\simeq\frac{d-2}{2(d-1)}\bigg(\frac{d}{2nB(\tfrac12,\tfrac1d)}\bigg)^{\frac d{d-2}}e^{-\frac{i\pi}{d-2}}\;x^{\frac{2(d-1)}{d-2}}\!,
\end{equation}
and in particular, in $d=4$,
\begin{equation}\label{eq:smallx4}
t_n^\pm\simeq2n\mathcal{Z}\mp\frac{4i}{3n^2B(\tfrac12,\tfrac14)^2}\,x^3,
\end{equation}
which is precisely the bouncing geodesic prediction \cite{Araya:2026shz}. 

While this analysis has been done for real $x$ under the assumption that appropriate contours of the $\omega$-integrals exist, the natural connection to WKB emerges when considering a complex $x$ as defined through $\mu$. Specifically, a choice of $\mu$ for which $\Im [\Psi(\mu)/\mu]=0$, there is an anti-Stokes line relating that value of $x$ directly to a turning point in the bulk $z_*$.  The $x\rightarrow 0$ limit then drives the bulk point to the curvature singularity.


\section{Discussion and generalisations}\label{s.Discusion}

In this work, we have focused on the planar AdS-Schwarzschild geometry. The computations presented here are, however, applicable to a broader context. In the most narrow sense, they are applicable to any bulk geometry which exhibits the same anti-Stokes topology as the cases studied here. On general grounds, this is expected for all settings where only one pair of QNM branches is present. This was exploited in \cite{Hartnoll:2026vhu}, where a relevant scalar deformation flows toward a Kasner interior. For example, in the case $d=3$ the power $\gamma$ is given by the Kasner exponent $p_t$, and the optical distance takes the form
\begin{equation}
    \gamma=\frac{1+p_t}{2(1-p_t)}, \quad \mathcal{Z}=\int_0^{(1+i\varepsilon) \infty} \sqrt{-\frac{g_{zz}}{g_{tt}}}\dd z.
\end{equation}
Our result \eqref{e.finaler} then precisely recovers the exponent $p$ for the boundary dual to such a spacetime, as was derived in \cite{Ceplak:2026ard,Oh:2026dky}.

More general settings will involve more QNM branches -- examples include charged black holes \cite{Ceplak:2025dds,AliAhmad:2026wem}, Gauss-Bonnet black holes \cite{futureHONG,Grozdanov:2018gfx}, etc. Nevertheless, our analysis is agnostic of possible other QNM branches as long as there is a branch associated with the anti-Stokes geometry described here. A more interesting and subtle question is about the connection between the exponent $\gamma$ and the presence or absence of a real bouncing geodesic. Our analysis breaks down if $\gamma \geq 1/2$, and it would be interesting to understand what happens to QNM scaling as well as the presence of the complex-time singularity in that case. This would be interesting to understand in charged black holes with Kasner interior \cite{Frenkel:2020ysx,Hartnoll:2020rwq} or regular black holes \cite{Chen:2026efa}. Another very interesting avenue would be to apply the same analysis for the mixmaster interior \cite{Belinsky:1970ew,Misner:1969hg,DeClerck:2023fax,Caceres:2026mug}.

Importantly, because the analysis relies only on QNMs, the singularity structure of $G(t,x)$ is in principle accessible by numerical means, even for bulk geometries which are not known in closed form (see e.g.~\cite{Konoplya:2017zwo,Jansen:2017oag}). This approach avoids trying to reach the singularity by means of crude numerical analytic continuation. 

Lastly, the bouncing singularity in complex time is expected to resolve away from the strict large-$N$ regime \cite{Dodelson:2025jff,Batra:2026gbq,Komargodski:2026oob,Buric:2026sit,Batra:2026otu,Dodelson:2026hjm,Almheiri:2026cpv,Chen:2026efa,Liu:2026luf}. It would be interesting to better understand this resolution in the Fourier picture, and translate the results to the $x\neq 0$ case.


\begin{acknowledgments}
S.V.~is funded by the Marie Skłodowska-Curie Actions programme GA-101177446 and the Slovenian Research and Innovation Agency (ARIS), contract number 5110-18/2025-5. M.V.~is funded by the NWA ORC programme Emergence at All Scales. In the process of creating this paper, LLMs were used exclusively as a resource and for error diagnosis. Calculations, plots, and text were all done by the authors and the authors alone.
\end{acknowledgments}


\bibliographystyle{apsrev4-1}
\bibliography{draft}

\end{document}